\documentclass[aps,pra,twocolumn,superscriptaddress,10pt,article,nofootinbib,showpacs,longbibliography]{revtex4-1}
\usepackage{amsfonts}
\usepackage{amsmath}
\usepackage{amsthm}
\usepackage{braket}
\usepackage{graphicx}
\usepackage{hyperref}
\usepackage{color}
\usepackage{amssymb}
\usepackage{dsfont}
\usepackage{verbatim}
\usepackage{amsmath,amscd}
\usepackage{multirow}
\usepackage{subfigure}

\newcommand{\mathscr}[1]{\ensuremath{\mathcal{#1}}}

\usepackage[normalem]{ulem}

\begin{document}

\title{Anomalous domain wall condensation in a modified Ising chain}

\author{Gertian Roose}
\affiliation{Department of Physics and Astronomy, University of Ghent, Krijgslaan 281, 9000 Gent, Belgium}
\author{Laurens Vanderstraeten}
\affiliation{Department of Physics and Astronomy, University of Ghent, Krijgslaan 281, 9000 Gent, Belgium}
\author{Jutho Haegeman}
\affiliation{Department of Physics and Astronomy, University of Ghent, Krijgslaan 281, 9000 Gent, Belgium}
\author{Nick Bultinck}
\affiliation{Department of Physics, University of California, Berkeley, California 94720, USA}

\begin{abstract}
We construct a one-dimensional local spin Hamiltonian with an intrinsically non-local, and therefore anomalous, global $\mathbb{Z}_2$ symmetry. The model is closely related to the quantum Ising model in a transverse magnetic field, and contains a parameter that can be tuned to spontaneously break the non-local $\mathbb{Z}_2$ symmetry. The Hamiltonian is constructed to capture the unconventional properties of the domain walls in the symmetry broken phase. Using uniform matrix product states, we obtain the phase diagram that results from condensing the domain walls. We find that the complete phase diagram includes a gapless phase that is separated from the ordered ferromagnetic phase by a Berezinskii-Kosterlitz-Thouless transition, and from the ordered antiferromagnetic phase by a first order phase transition.
\end{abstract}

\maketitle

\section{Introduction}

Spontaneous symmetry breaking in quantum many-body systems can be characterized by the non-zero expectation value of an order parameter. In symmetry broken systems there exists a basis such that each ground state is uniquely characterized by its uniform and non-zero value for the order parameter. For certain symmetry breaking patterns and in certain spatial dimensions it is possible to consider states where the order parameter is non-uniform and contains a topological defect \cite{Mermin,Coleman}, such as for example a domain wall in one dimension or a vortex in two dimensions. Because such topological defects are stable and cannot be created by local operators, it is possible that they bind fractional quantum numbers associated with unbroken global symmetries. In fact, many examples of systems where this occurs are known. Among the earliest examples are the Jackiw-Rebbi \cite{JackiwRebbi} or the Su-Schrieffer-Heeger model \cite{SSH}, where domain walls bind half-integer U$(1)$ charge, and the spin-$1/2$ soliton in quantum spin chains \cite{Fadeev}.

The binding of fractional quantum numbers to topological defects is closely related to Lieb-Schultz-Mattis-Oshikawa-Hastings (LSMOH) theorems \cite{LSM,Oshikawa1,Hastings}, which forbid the existence of short-range entangled phases that do not break any microscopic on-site and spatial symmetries. In the original LSMOH theorems, the relevant on-site symmetry was U$(1)$ or SU$(2)$, and the spatial symmetry was simply lattice translation symmetry. However, by now LSMOH theorems exist for many other on-site and spatial symmetries \cite{Parameswaran,zaletel,Watanabe,Po,Oshikawa3}. An intuitive way to understand the connection between LSMOH theorems and fractionalization of topological defects is to imagine a system in a symmetry broken phase, such that condensation of topological defects drives it to a disordered phase. If the defects carry fractional quantum numbers, this condensation transition cannot result in a short-range entangled, featureless state. An interesting example is the spin-$1/2$ Heisenberg antiferromagnet on the square lattice. Because this model has half-odd integer spin per unit cell, the LSMOH theorem forbids a short-range entangled symmetric ground state. If we assume the ground state is a valence-bond solid (VBS), topological defects in the $\mathbb{Z}_4$ valued order parameter carry spin-$1/2$ \cite{LevinSenthil}. In the N\'{e}el phase, spin-wave excitations in topological sectors with an odd skyrmion number carry momenta around $(0,\pi)$ or $(\pi,0)$ \footnote{Note that these momenta are not fractionalized. Only for on-site symmetries does the quantum number have to be fractional in order for the defect to be non-trivial, because the integer part can be changed by adding local excitations.} \cite{Haldane}. Condensing the skyrmions therefore leads to the fourfold ground state degeneracy of the VBS phase  \cite{Read,ReadSachdev}. These observations also lie at the basis for the original theory of deconfined quantum criticality \cite{DQCP}, which was proposed to describe a Landau-forbidden continuous phase transition between the N\'{e}el and VBS orders. Recently, a 1D Hamiltonian with a LSMOH constraint was constructed such that in the VBS phase the domain walls bind a non-trivial projective representation of $\mathbb{Z}_2\times\mathbb{Z}_2$ \cite{Jiang}. The authors of Ref. \cite{Jiang} argued that condensing these domain walls results in a deconfined quantum critical point in the phase diagram, separating two different symmetry-broken phases.

Another context in which fractionalized quantum numbers and/or unconventional zero modes bind to topological defects, is at the boundary of symmetry-protected topological (SPT) phases \cite{FidkowskiKitaev,PollmannBerg,1Dclassification,CZX,ChenWen}. One of the most notable examples is the time-reversal symmetric superconducting boundary state of the 3D topological insulator, where a vortex traps a Majorana mode \cite{FuKane}. In the non-trivial 3D bosonic SPT phase with U$(1)\times \mathbb{Z}_2^T$ symmetry, where $\mathbb{Z}_2^T$ is time-reversal symmetry, boundary vortices bind a Kramers pair in their core \cite{VishwanathSenthil}. In Ref. \cite{JuvenWang}, the authors showed that for certain two-dimensional SPT phases with $\mathbb{Z}_n\times\mathbb{Z}_m$ symmetry, boundary domain walls associated with broken $\mathbb{Z}_n$ symmetry carry fractional charge under $\mathbb{Z}_m$, and vice versa. When the symmetry group is $\mathbb{Z}_n\times\mathbb{Z}_m\times\mathbb{Z}_p$, there exist 2D bosonic SPT phases such that a boundary domain wall of $\mathbb{Z}_n$ binds a non-trivial projective representation of $\mathbb{Z}_m\times\mathbb{Z}_p$ \cite{JuvenWang}. 

It was recognized early on that the physics of deconfined quantum critical points and the boundaries of SPT phases are closely related \cite{VishwanathSenthil}. More recently, systems where a LSMOH theorem applies were interpreted as the boundary of a SPT phase with both on-site and spatial symmetries \cite{surface,jian}. All these systems also share the property that a topological theta term or Wess-Zumino-Witten term is essential to obtain the correct non-linear sigma model effective field theory \cite{Haldane,Tanaka,Senthil,CenkeXu,jian}. The physical meaning of such terms is exactly that they provide the topological defects with the correct properties such as fractional quantum numbers. By now, a deeper unified language for the physics of LSMOH theorems, SPT surface states and deconfined quantum critical points has emerged in terms of `t Hooft anomalies \cite{Ryu2,Wen,Kapustin,Ryu,MetlitskiThorngren,Furuya,Nahum}. For UV lattice models, a `t Hooft anomaly simply means that a global symmetry is realized in an intrinsically non-local way \cite{CZX,Wen}. In the context of LSMOH theorems, `t Hooft anomalies can occur because of the non-on-site nature of the spatial symmetries. In the context of SPT phases, a local symmetry in the bulk can effectively act as a non-local symmetry on the low-energy boundary or surface modes. The common wisdom is that when a non-local symmetry with non-trivial `t Hooft anomaly gets spontaneously broken, topological defects in the corresponding order parameter will acquire unconventional properties such as fractional quantum numbers. However, it is important to note that fractionalization is not the only unconventional property of defects that can occur when a non-local symmetry gets broken. Another possibility is that the defects have non-trivial statistics \cite{Wilczek}. For example, on the boundary of a 3D bosonic topological insulator with U$(1)\rtimes \mathbb{Z}^T_2$ symmetry the vortices become fermions \cite{VishwanathSenthil,Metlitski}. 

In this work, we consider a 1D model with a $\mathbb{Z}_2$ `t Hooft anomaly where similar phenomena occur. In particular, we construct a spin Hamiltonian for which in the symmetry-broken phase the domains walls between the two vacua behave as semions in a sense that we specify below. It can be interpreted as the edge Hamiltonian of a 2D bosonic SPT phase corresponding to the non-trivial element of $H^3(\mathbb{Z}_2,U(1))=\mathbb{Z}_2$ \cite{CZX,LevinGu,ChenWen}. The Hamiltonian is closely related to the 1D quantum Ising model in transverse magnetic field, and contains a parameter that we can tune to condense the domain walls. We show that this model indeed has a non-local $\mathbb{Z}_2$ symmetry that can be written in matrix product operator form. We find that the Hamiltonian we construct has close connections to anyon chains \cite{Trebst,Trebst2}, and that --not surprisingly-- its symmetry is of the CZX-type \cite{CZX}. Upon condensing the `semionic' domain walls, there occurs a Berezinskii-Kosterlitz-Thouless (BKT) transition \cite{Berezinski,KT} to a Luttinger liquid phase \cite{LuttingerLiquid} with an emergent U$(1)\times$U$(1)$ symmetry. The Luttinger liquid description of the gapless regime agrees with the Chern-Simons description of 2D SPT phases \cite{LuVishwanath}. We numerically study the entire phase diagram with uniform matrix product states and also find a first order transition, making the phase diagram very similar to that of the XXZ model. We expect our model to capture the generic boundary phase diagram of the non-trivial 2D bosonic $\mathbb{Z}_2$ SPT phase. In Ref. \cite{WangWenWitten}, the authors constructed gapped boundaries of SPT phases using symmetry extentions. However, we did not find a natural way to incorporate these symmetry extensions in our minimal effective model for the domain walls.

\section{`Semionic' domain walls}

We imagine a situation where the $\mathbb{Z}_2$ symmetry of a 2D non-trivial bosonic SPT is spontaneously broken on the boundary. We also assume that the symmetry-breaking induced gap is much smaller than the bulk gap. In this case, the low-energy degrees of freedom will be the domain walls on the boundary and the dynamics will be effectively one-dimensional. In this section we discuss the imprint of the $\mathbb{Z}_2$ `t Hooft anomaly on the boundary symmetry-breaking phase. 

Based on the group cohomology classification of bosonic SPT phases \cite{ChenWen}, a natural guess for the property of the boundary domain walls that distinguishes them from conventional Ising domain walls, is that they have unusual fusion rules. In particular, if we consider three domain walls localized in some region, pair-wise annihilating the first two or pair-wise annihilating the last two gives a relative minus sign. Schematically,
\begin{equation}\label{fmove}
(1,2)3=-1(2,3)\,,
\end{equation}
where we numbered the domain walls and the brackets denote a pairwise annihilation process. That this is indeed the correct property of the anomalous domain walls can be verified by the tensor network constructions of 2D SPT phases \cite{CZX,Williamson2}. The intuition behind these unusual fusion rules is now, analogous to Haldane's argument for the gaplessness of the spin-1/2 chain \cite{HaldaneGap2,HaldaneGap,Haldane}, that in a path integral representation these minus signs will lead to destructive interference which prevents the disordered phase from having short-range correlations.

Let us elaborate on why Eq. \eqref{fmove} implies that the domain walls can be interpreted as semionic quasiparticles. For this we consider a state with $2N$ domain walls. We order the domain walls and pair them up in neighbouring pairs, i.e. we represent the state as
\begin{equation}
(1,2)(3,4)\dots (2N-1, 2N)\, ,
\end{equation}
where now we interpret the brackets as indicating that these domain wall pairs were created together from the vacuum. This choice of pairing is arbitrary and merely serves as a reference configuration. Let us now create an additional domain wall pair. There are two possibilities to do this. The first is that we create the pair between two other pairs, such that the state e.g. becomes
\begin{equation}
(1,2)(3,4)(1',2')(5,6)\dots (2N-1,\, 2N)\, ,
\end{equation}
where we denote the newly created pair with primes. This state can simply be relabeled to obtain the reference state with $2N+2$ domain walls:
\begin{multline}
(1,2)(3,4)(1',2')(5,6)\dots (N-1,\, N)\\
 \rightarrow (1,2)(3,4)(5,6)(7,8)\dots(2N+1,\, 2N+2)\, .
\end{multline}
The second possibility is that we create the pair in between two domain walls that were paired up in the reference state. In that case we obtain for example
\begin{equation}
(1,2)(3,4)(5(1',2')6)\dots (2N-1,\, 2N)\, .
\end{equation}
Now applying rule \eqref{fmove} implies that this state is equal to
\begin{multline}
(1,2)(3,4)(5(1',2')6)\dots (2N-1\,, 2N) \\
\rightarrow - (1,2)(3,4)(5,6)(7,8)\dots (2N+1,\, 2N+2)\, .
\end{multline}
So we see that the creation of a pair of domain walls at position $x$ gives a minus sign if the number of domain walls to the left of $x$ is odd, while we get no minus sign if the number of domain walls to the left of $x$ is even. This implies that the creation operator for a \emph{single} domain wall at site $x$ would produce a factor of $\pm i$ if the number of domain walls to the left is odd, which agrees with the findings of Ref. \cite{LuLee}. This justifies the term semionic, which refers to `half-fermion' statistics.

\section{Effective model}\label{sec:ham}

\subsection{The Hamiltonian}

In this section we construct a 1D effective Hamiltonian that captures the semionic nature of the domain walls described above. As a first step, we recall the Kramers-Wannier self-duality mapping of the 1D quantum Ising model in a transverse magnetic field. If we call the original Ising spins the $\sigma$-spins, then we can introduce $\tau$-spins living in between the Ising spins which represent domain walls. We use the convention that a $\tau$-spin is zero if its two neighbouring $\sigma$-spins are equal and is one when its neighbouring spins are different. The Ising Hamiltonian can be written either in terms of the original $\sigma$-spins, or in terms of the domain walls represented by $\tau$-spins:
\begin{eqnarray}
H&=&\sum_i -J \sigma^z_i\sigma^z_{i+1} + B\sigma^x_i \nonumber \\
 \leftrightarrow H'&=& \sum_i -J \tau^z_{i+1/2} + B \tau^x_{i-1/2}\tau^x_{i+1/2}\, .
\end{eqnarray}
Here $\tau^{i}$ represent the Pauli matrices, but acting on the domain wall states. The $\sigma$-spins are taken to live on the integer-valued lattice sites, while the $\tau$-spins live on the half-integer lattice sites. Under this duality the ferromagnetic interaction of the $\sigma$-spins maps to a chemical potential for the $\tau$ domain walls, while the magnetic field maps to a hopping and pair creation term for the domain walls. This is easily understood, since increasing the ferromagnetic interacting suppresses the existence of domain walls while the magnetic field will flip $\sigma$-spins and cause domain walls to be created and move around. So $H'$ represents the dynamics of domain walls, which can condense (by lowering their chemical potential) and give rise to a disordered phase. 

\begin{center}
\begin{figure}
a) \hspace{0.1 cm}
\includegraphics[width=0.3\textwidth]{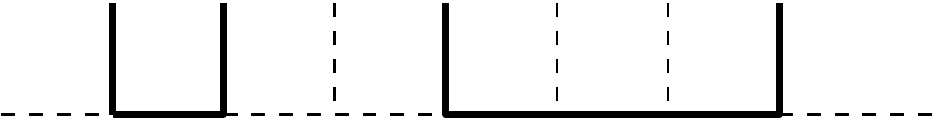}\\
\vspace{0.5 cm}
b)  \hspace{0.1 cm}
\includegraphics[width=0.27\textwidth]{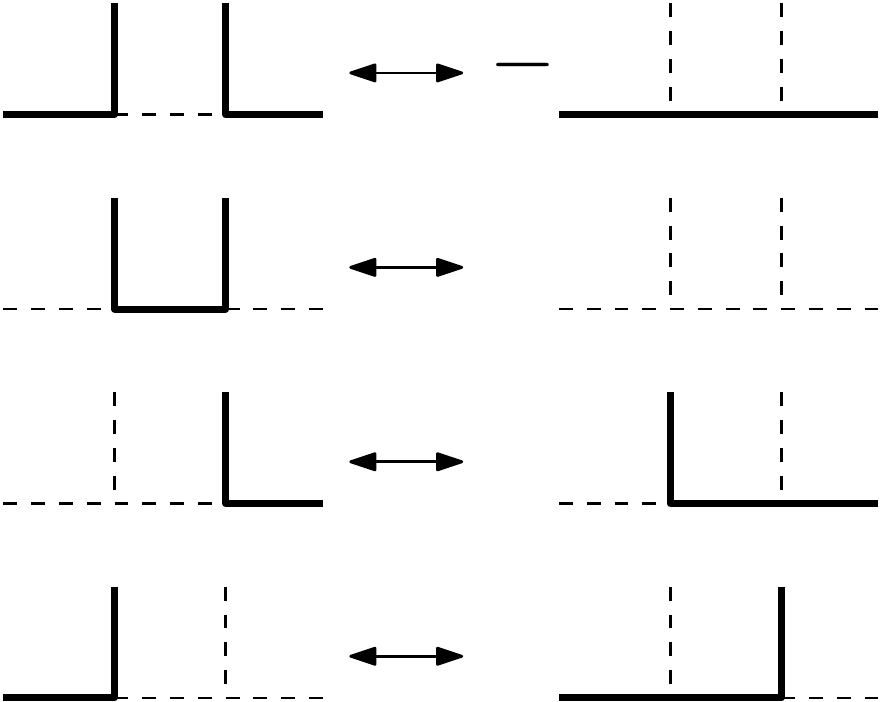}
\caption{a) Graphical representation of the state $|0\tilde{1}1\tilde{1}0\tilde{0}0\tilde{1}1\tilde{0}1\tilde{0}1\tilde{1}0\rangle$, where $\tilde{j}$ denote the $\tau$-spins and $j$ denote the $\sigma$-spins. b) Action of the Hamiltonian term $H_{Dyn}$ on states in the subspace satisfying $\sigma^z_{i}\tau^z_{i+1/2} \sigma^z_{i+1} = 1$.}\label{fig:hilb}
\end{figure}
\end{center}

To construct a Hamiltonian that has semionic domain walls we first keep both the $\sigma$ and $\tau$ spins. Since now our Hilbert space consists of both the $\sigma$ and $\tau$ spins simultaneously, we need a term that enforces the $\tau$-spins to represent domain walls of the $\sigma$-spins. This is easily done with a $\mathbb{Z}_2$ Gauss-law term:
\begin{equation}
H_{Gauss} = -g\sum_i \sigma^z_{i}\tau^z_{i+1/2} \sigma^z_{i+1}\, .
\end{equation}
The Gauss constraint term commutes with all terms that we will add to the Hamiltonian later on, so by taking $g>0$ large enough the low-energy states will live in the subspace where the $\tau$-spins represent domain walls of the $\sigma$-spins. This subspace characterized by $\sigma^z_{i}\tau^z_{i+1/2} \sigma^z_{i+1} = 1$ has a nice graphical representation. If we represent the $\sigma$-spins as horizontal links and the $\tau$-spins as vertical links, such that we obtain a one-dimensional lattice that is a sequence of coordination number three vertices, then the low-energy subspace is in one-to-one correspondence with all coverings of this lattice with strings that end on the vertical links. The precise correspondence is that a $0$-state represents the absence of a string, and the $1$-state represents the presence of a string. We illustrate this graphical representation in Fig. \ref{fig:hilb}(a).

The chemical potential term for the domain walls is the same as in the Ising model:
\begin{equation}
H_{\mu} =- \mu \sum_i \tau^z_{i+1/2}\, .
\end{equation}
As a final term, we need an analogue of the domain wall hopping/pair creation term $\tau^x_{i-1/2}\tau^x_{i+1/2}$ of the transverse field Ising model. Denoting the domain wall states with $\{|\tilde{0}\rangle\, ,|\tilde{1}\rangle\}$ and the $\sigma$-spin states with $\{|0\rangle\, ,|1\rangle\}$, we define the term $H_{Dyn}$ by its action on any subsequent $\tau$-$\sigma$-$\tau$ triplet as
\begin{eqnarray}
|\tilde{1}0\tilde{1}\rangle &\leftrightarrow & - |\tilde{0}1\tilde{0}\rangle \label{creation1}\\
|\tilde{1}1\tilde{1}\rangle &\leftrightarrow & |\tilde{0}0\tilde{0}\rangle \label{creation2}\\
|\tilde{0}0\tilde{1}\rangle &\leftrightarrow & |\tilde{1}1\tilde{0}\rangle \label{hopping1} \\
|\tilde{1}0\tilde{0}\rangle &\leftrightarrow & |\tilde{0}1\tilde{1}\rangle \label{hopping2}\, ,
\end{eqnarray}
and $H_{Dyn}$ is zero on any state that violates the Gauss-term. We give a graphical representation of the action of $H_{Dyn}$ in Fig. \ref{fig:hilb}(b). The only difference between $H_{Dyn}$ and the conventional Ising model term $\tau^x_{i-1/2}\tau^x_{i+1/2}$ is the minus sign in \eqref{creation1}. This term represents the creation of a domain wall pair when there are an odd number of domain walls to the left of it. Equation \eqref{creation2} also represents a pair creation/annihilation process, but with an even number of domain walls to the left. The easiest way to see this is to look at Fig. \ref{fig:hilb}(b), and to realize that the $\sigma$-spins encode the parity of the number of $\tau$-spins to the left of it. Equations \eqref{hopping1} and \eqref{hopping2} represent domain wall hopping. 

We now take the Hamiltonian $H$ to be the sum of all preceding terms:
\begin{equation}\label{totalham}
H=H_{Gauss}+H_\mu + H_{Dyn}\, .
\end{equation}
We claim that this Hamiltonian captures the universal low-energy physics at the boundary of the non-trivial 2D bosonic $\mathbb{Z}_2$ SPT phase. In the next section we first discuss the global $\mathbb{Z}_2$ symmetry of this Hamiltonian. In section \ref{sec:num} we numerically study the phase diagram of $H$ as a function of $\mu$ and show that this model indeed does not have a gapped, disordered phase, which is the hallmark of the edge physics of a non-trivial 2D SPT phase. At this point we also want to mention that for $\mu=0$, our Hamiltonian $H$ is very closely related to the anyonic chains that have previously been constructed in the literature \cite{Trebst,Trebst2}. Specifically, our model at $\mu=0$ can be obtained by constructing an anyon chain with the $F$-symbols of the SU$(2)_1$ modular category. However, there is one important difference compared to the usual anyonic chain construction. In Refs. \cite{Trebst,Trebst2} the vertical links, corresponding to our domain wall or $\tau$ degrees of freedom, are fixed while here they are allowed to fluctuate. We will see below that the connection with the SU$(2)_1$ anyon chain at $\mu=0$ fits nicely with the phase diagram we obtain numerically. 

\subsection{$\mathbb{Z}_2$ symmetry}
 
Although the Hamiltonian in Eq. \eqref{totalham} is very closely related to the transverse Ising model, it does not have the same $\mathbb{Z}_2$ symmetry $\bigotimes_{i}\sigma^x_i$. However, $H$ does have a \emph{low-energy} $\mathbb{Z}_2$ symmetry. To expose it, we focus on states in the Hilbert space that do not violate the Gauss term, i.e. we only consider states that satisfy $\sigma^z_{i}\tau^z_{i+1/2} \sigma^z_{i+1} = 1$. We now claim that the relevant $\mathbb{Z}_2$ symmetry is given by
\begin{equation}\label{Z2}
(-1)^{\# \text{strings}}\bigotimes_{i} \sigma_{i}^x\, ,
\end{equation}
i.e. it flips all the $\sigma$-spins and adds a minus sign when the number of strings is odd. Note that we can add the minus sign before or after flipping all the $\sigma$-spins, since this does not change the number of strings. Because the minus sign commutes with the product of $\sigma^x$, it is clear that this symmetry squares to the identity. 

\begin{center}
\begin{figure}
\includegraphics[width=0.35\textwidth]{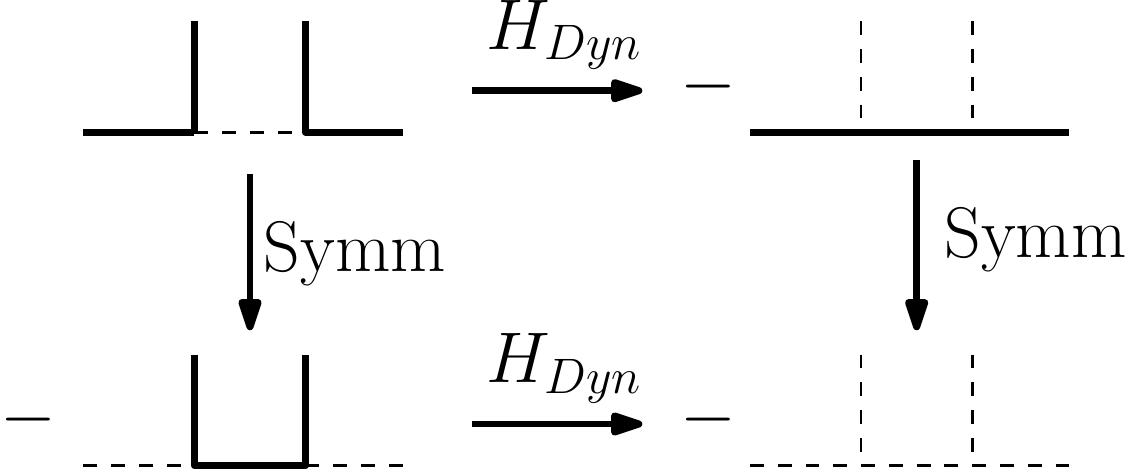}
\caption{Graphical illustration of the commutation relation between the $\mathbb{Z}_2$ symmetry in Eq. \eqref{CZx} and $H_{Dyn}$. Both paths in the diagram (first $H_{Dyn}$, then the symmetry and vice versa) commute.}\label{fig:commute}
\end{figure}
\end{center}

The sign $(-1)^{\#\text{strings}}$ appears to be a very non-local operator. However, we can encode it via local operators by noting that counting the number of strings is equivalent to counting the number of right-hand endpoints of strings. A right-hand endpoint of a string can be detected locally, and we can assign a minus sign to it using a diagonal matrix for every $\tau$-spin and the $\sigma$-spin to the left of it. We then let the operator add a minus sign when both these spins are one, which indeed corresponds to the situation where a string comes from the left and ends on that $\tau$-spin. Concretely, if we define
\begin{equation}\label{CZ}
CZ_{i} = |0\tilde{0}\rangle\langle0\tilde{0}| + |1\tilde{0}\rangle\langle1\tilde{0}| + |0\tilde{1}\rangle\langle0\tilde{1}| - |1\tilde{1}\rangle\langle1\tilde{1}|
\end{equation}
to act on $\sigma$-spin $i$ and $\tau$-spin $i+1/2$, then the symmetry can be written as a product of local matrices as
\begin{equation}\label{CZx}
\bigotimes_i CZ_i \bigotimes_i \sigma^x_{i}.
\end{equation}
Using the graphical representation one can easily check that this operator commutes with $H_{Dyn}$, as we illustrate in Fig. \ref{fig:commute}. The operator \eqref{CZx} also trivially commutes with $H_\mu$.

There are a few important points we need to clarify about the $\mathbb{Z}_2$ symmetry. The operators $\bigotimes_i CZ_i$ and $\bigotimes_i \sigma^x_{i}$ do not commute. As a result, the operator in equation \eqref{CZx} does not square to the identity and does not truly represent a $\mathbb{Z}_2$ symmetry. It is only when we project into the low-energy subspace satisfying the Gauss constraint $\sigma^z_{i}\tau^z_{i+1/2} \sigma^z_{i+1} = 1$ that it acts as in equation \eqref{Z2}, and is a true $\mathbb{Z}_2$ symmetry. If we take the tensor product of local matrices in equation \eqref{CZx} and project it into the subspace satisfying the Gauss term, then we end up with a non-local matrix product operator (MPO) representation of the $\mathbb{Z}_2$ symmetry. So the correct statement is that the Hamiltonian we are considering has a $\mathbb{Z}_2$ MPO symmetry in its low-energy subspace satisfying the Gauss term.

\subsection{Simplified Hamiltonian}

The Hamiltonian in Eq. \eqref{totalham} has the clearest physical interpretation in terms of the dual domain wall variables we used in the previous sections. However, we can also reformulate it using only the original $\sigma$ spins. In these variables, the Hamiltonian \eqref{totalham} becomes
\begin{equation}\label{Hamls}
H=\sum_i CZ_{i-1,i+1}\sigma^x_i  -\mu \sigma^z_i\sigma^z_{i+1}\, , 
\end{equation}
where we introduced the notation $CZ_{ij}$, which is the same matrix as defined in Eq. \eqref{CZ}, but now acting on the $\sigma$-spins labeled by $i$ and $j$. Since we have omitted the domain walls, we no longer need the Gauss term in the Hamiltonian. In terms of the $\sigma$-spins, the $\mathbb{Z}_2$ symmetry becomes
\begin{equation} \label{eq:mpo}
\bigotimes_i \sigma^x_i \bigotimes_i CZ_{i,i+1}\sigma^z_i \,.
\end{equation}
One can check that this operator, which is now manifestly a MPO, indeed squares to the identity and commutes with the Hamiltonian \eqref{Hamls}. Because this MPO corresponds to the non-trivial element in $H^3(\mathbb{Z},\mathrm{U}(1))$, the Hamiltonian in Eq. \eqref{Hamls} cannot have a unique, short-range entangled ground state \cite{CZX}.

\section{Condensation of domain walls: field theory analysis}

Before discussing our numerical results in the next section, we first turn to a low-energy field theory analysis. Given that gapless boundary modes of two-dimensional bosonic symmetry-protected phases are known to have a Luttinger liquid description \cite{LuVishwanath}, we expect that if our model has a gapless regime it will flow to this effecive field theory in the IR. Here we follow the conventional notation (with units such that $u=1$) \cite{Giamarchi}, and write the Luttinger liquid or compact free boson action as
\begin{eqnarray}
\mathcal{L}_0 & = &\frac{1}{2\pi K}(\partial_\mu\phi)^2\, .
\end{eqnarray}
The compactification radius of $\phi$ is taken to be $\pi$. We define the dual field $\theta(x)$ via the relation $\partial_x\theta = \partial_t\phi/K$. Canonical quantization implies that the boson fields $\phi$ and $\theta$ satisfy the commutation relation
\begin{equation}
[\phi(x),\partial_y\theta(y)]=i\pi\delta(x-y)\, .
\end{equation}
From the canonical commutation relation it follows that the operator which shifts $\phi$ by a constant $\alpha$ is given by $\exp\left(-i\frac{\alpha}{\pi}\int\mathrm{d}x\,\partial_x\theta(x)\right)$. The compactification condition on $\phi$ implies that this operator should be the identity operator when $\alpha = \pi$, which implies that $\theta$ is also compact with compactification radius $2\pi$. We can write the Luttinger liquid Hamiltonian as
\begin{equation}\label{ham}
H=\frac{1}{2\pi}\int\mathrm{d}x\, \left(K(\partial_x\theta)^2+\frac{1}{K}(\partial_x\phi)^2 \right)\, .
\end{equation}
From this Hamiltonian we recognize the R-duality $2\phi\leftrightarrow \theta$, $K\leftrightarrow 1/4K$ of the free boson CFT. 

As was shown in previous works \cite{Chen,LevinGu,LuVishwanath}, in the non-trivial SPT phase the global $\mathbb{Z}_2$ symmetry acts on the fields as
\begin{equation}\label{symmaction}
\phi\rightarrow \phi+\frac{\pi}{2}\;,\hspace{0.5 cm} \theta\rightarrow \theta+\pi\, .
\end{equation}
From this symmetry action, we see that there exist no $\mathbb{Z}_2$ symmetric terms of the form $\cos(2m\phi)$ or $\cos(m\theta)$ with $m\in\mathbb{Z}$ that we can add to the Luttinger liquid Lagrangian to create a gap, and at the same time obtain a unique, symmetric ground state. This is the fingerprint of the `t Hooft anomaly, which excludes the existence of a gapped, disordered phase. 

The global symmetry operator which implements the shifts in Eq. \eqref{symmaction} is given by $\exp\left(-i\int\mathrm{d}x \left[\frac{1}{2}\partial_x\theta(x)+\partial_x\phi(x) \right]\right)$. It follows that the operator which creates a domain wall at position $x$ is given by \cite{LuLee}

\begin{equation}
\hat{D}^\dagger(x) = e^{-i\left(\frac{1}{2}\theta(x)+\phi(x) \right)}
\end{equation}
At this point, we import a result from our numerical simulations presented in the next section. As detailed below, we find that the translation symmetry of the lattice Hamiltonian acts an internal $\mathbb{Z}_3$ symmetry in the low-energy Luttinger liquid description. Specifically, our numerics show that under a translation by one lattice site, the domain wall creation operator $\hat{D}^\dagger$ picks up a phase $e^{i2\pi/3}$. This is to be compared with the Luttinger liquid description of the XXZ spin chain, where translation symmetry acts as an internal $\mathbb{Z}_2$ symmetry in the effective field theory. The $\mathbb{Z}_3$ symmetry action on the domain wall creation operator does not allow us to unambiguously determine its action on the boson fields $\phi$ and $\theta$. However, we do not expect the $\mathbb{Z}_3$ symmetry to be anomalous because the gapped ferromagnetic phase of our lattice Hamiltonian is translationally invariant. So we can without loss of generality take the internal $\mathbb{Z}_3$ symmetry to act as \cite{LuVishwanath}

\begin{equation}
\phi\rightarrow \phi\, ,\hspace{0.5 cm} \theta\rightarrow \theta - \frac{4\pi}{3}\, .
\end{equation}

The scaling dimension of $\cos(m\phi)$ is $\frac{m^2K}{4}$, while the scaling dimension of $\cos(m\theta)$ is given by $\frac{m^2}{4K}$. The most RG-relevant perturbations of the Luttinger liquid respecting all symmetries are therefore given by $\cos(4\phi)$ and $\cos(6\theta)$. So in the parameter range $1/2< K < 9/2$, the Luttinger liquid has no symmetry respecting relevant perturbations. The $\cos(4\phi)$ term is irrelevant for $K>1/2$, while it is relevant for $K<1/2$. So at $K=1/2$, which is the self-dual point of the free boson CFT, there is a BKT transition to a gapped phase where the $\phi$ field gets pinned to one of the minima of the $\cos(4\phi)$ term. This gapped phase spontaneously breaks the global $\mathbb{Z}_2$ symmetry, but preserves translation symmetry. We therefore identify it with the ferromagnetic phase of our microscopic Hamiltonian [Eq. \eqref{Hamls}] obtained for large $\mu>0$. From the perspective of the ferromagnetic phase, the BKT transition into the Luttinger liquid phase results from the condensation of $\mathbb{Z}_2$ domain walls. As noted above, the Hamiltonian in Eq. \eqref{Hamls} constructed to describe this domain wall condensation takes the form of an SU$(2)_1$ anyon chain when $\mu=0$. Therefore, it is natural to expect that the BKT transition in this model will occur at $\mu=0$, since the self-dual point of the free boson CFT is equivalent to the SU$(2)_1$ Wess-Zumino-Witten CFT. We will confirm this in the next section containing our numerical results.

The sine-Gordon Lagrangian $\mathcal{L}_0+g\cos(4\phi)$ of course also describes the original BKT transition in the 2D classical XY model~\cite{Nelson} or 1D quantum XXZ Hamiltonian. However, there is one important distinction compared to the present discussion. In the XY model, there is a microscopic U$(1)$ symmetry on both sides of the BKT transition, which in the field theory language is associated with the charge $Q=\int \mathrm{d}x\,\partial_x\phi$, i.e. the winding of the boson field. In the Luttinger liquid phase, there is an additional emergent U$(1)$ symmetry, with charge $\tilde{Q}=\int\mathrm{d}x\, \partial_t\phi$. In the semionic domain wall Hamiltonian, both U$(1)$ symmetries are emergent and only present in the Luttinger liquid phase. This distinction does not appear in the field theory description however, which captures the behavior around the Luttinger liquid fixed point. Once the cosine term becomes relevant, the theory will flow to a gapped fixed point, where the $\mathrm{U}(1)$ symmetry ceases to have any physical meaning.

\section{Numerical results}\label{sec:num}

\begin{figure}[t] \centering
\includegraphics[width=0.8\columnwidth]{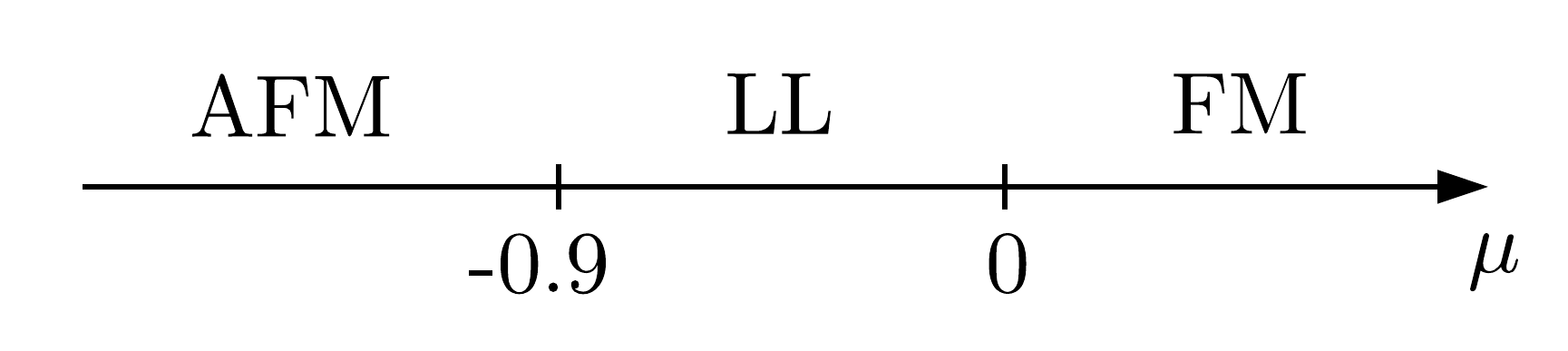}
\caption{The phase diagram of the Hamiltonian [Eq.~\eqref{totalham}] as obtained by uniform MPS simulations. Both the ferromagnetic (FM) and antiferromagnetic (AFM) phases can be understood in the limit of large $|\mu|$ by the relation to the standard transverse-field Ising model. The semionic nature of the domain-wall hopping and creation introduces a gapless Luttinger liquid (LL) phase in between; the transitions are of the BKT type (LL $\to$ FM) and first-order (LL $\to$ AFM).}
\label{fig:diagram}
\end{figure}

In this section, we explore the phase diagram of our effective model numerically, and confirm the theoretical considerations above. Our simulations were performed using tangent-space methods for uniform matrix product states (MPS) \cite{Vanderstraeten2018}; in particular, we use the vumps algorithm \cite{Zauner-Stauber2018} for finding variational MPS approximations for the ground state of the Hamiltonian, and apply the quasiparticle excitation ansatz \cite{Haegeman2012a} for computing the low-lying excited states. Because the framework of uniform MPS works directly in the thermodynamic limit, we do not experience any finite-size errors, and the only refinement parameter is the MPS bond dimension. For simplicity, we simulate the model using the reduced form of the Hamiltonian in Eq.~\eqref{Hamls}. In Fig.~\ref{fig:diagram} we summarize the phase diagram that we have obtained by our simulations.

\par
\begin{figure}[t] \centering
\includegraphics[width=0.99\columnwidth]{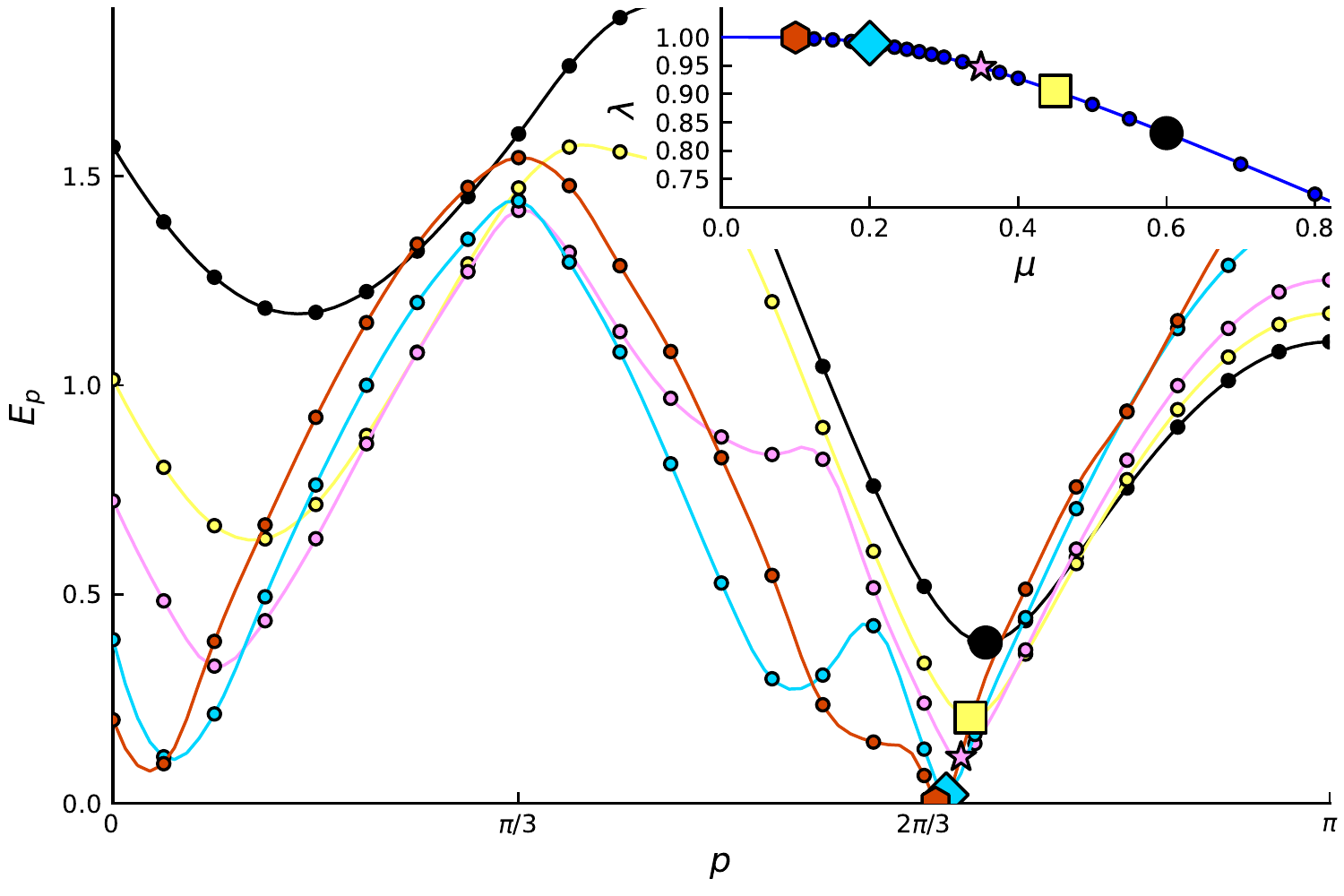}
\caption{The excitation energy as a function of momentum in the topological (domain wall) sector for five values of $\mu$ (the spectrum is reflection symmetric for $p\to -p$). Energies were obtained with the MPS quasiparticle ansatz with an MPO string and bond dimensions up to $D$ = 100. We observe that the minimum of the dispersion relation (indicated with the larger symbol) shifts as $\mu$ varies, where the gap closes at momentum $\pm 2\pi/3$ as $\mu=0$ is approached. In the inset we provide the value of $\lambda=\bra{\psi_\mathrm{MPS}}  O \ket{\psi_\mathrm{MPS}}^{1/N}$ as a function of $\mu$ (for system size $N\to\infty$), and we indicate the five points for which we have computed the spectrum.}
\label{fig:fm}
\end{figure}

 We start in the ferromagnetic phase ($\mu>0$). In the limit of large $\mu$ we recover the standard ferromagnetic transverse-field Ising model, for which the order parameter $\braket{\sigma^z_i}$ signals the $\mathbb{Z}_2$ symmetry breaking. For the Ising model, a variational MPS simulation generically yields one of the two states with maximal symmetry breaking; moreover, since these two ground states are connected by the symmetry operation $\bigotimes_i\sigma^x_i$, these two ground states have exactly the same entanglement structure. For the non-local MPO symmetry [Eq.~\eqref{eq:mpo}] in our model, this is no longer the case as we always find the same MPS $\ket{\psi_\mathrm{MPS}}$ as a variational optimum at a given bond dimension. The second ground state is found by acting with the MPO on the first, which increases the bond dimension. Correspondingly, the entanglement spectra of the two ground states are different and, in particular, the bipartite entanglement entropy of $\ket{\psi_\mathrm{MPS}}$ is smaller. As MPS ground-state approximations induce a bias towards low-entanglement states, this explains why we find only one variationally optimal ground state at a given bond dimension. To characterize the $\mathbb{Z}_2$ symmetry breaking in our simulations we compute the quantity $\lambda = \bra{\psi_\mathrm{MPS}}  O \ket{\psi_\mathrm{MPS}}^{1/N} $, where $O$ is the MPO operator in Eq.~\eqref{eq:mpo} and $N$ is the diverging system size; in uniform MPS simulations this `overlap per site' is easily computed directly in the thermodynamic limit. From the inset of Fig.~\ref{fig:fm} we clearly see that the $\mathbb{Z}_2$ symmetry is spontaneously broken in the ferromagnetic phase, but that the symmetry breaking vanishes as $\mu=0$ is approached.

\par In a system where different ground states are related through an MPO symmetry, the elementary excitations have a topological nature, in the sense that they are created by a local operator with an MPO string attached \cite{Marien2017}. Here, the MPO string serves as the generalization of the Jordan-Wigner string in the Ising model. The MPS quasiparticle ansatz is straightforwardly generalized to the case of MPO strings \cite{Marien2017}, and we can compute the excitation energy within the non-trivial topological sector for every value of the momentum. In Fig.\ref{fig:fm} we plot the spectrum for different values of $\mu$. Interestingly, the absolute minimum of the dispersion relation continuously shifts from momentum $p=\pi$ in the Ising limit to $p=\frac{2\pi}{3}$ at the critical point where the gap closes. This tells us that the domain wall creation operator in a long-wavelength continuum theory for the gapless phase picks up a phase $e^{i2\pi/3}$ under translation. So in an effective field theory description, translation symmetry will act as an internal $\mathbb{Z}_3$ symmetry. In the previous section we used this result in our Luttinger liquid analysis. Next to the gap closing at $p=\frac{2\pi}{3}$, we also find additional local minima close to momentum zero and $\frac{2\pi}{3}$, that correspond to the lower edges of the three and five particle continuum respectively (note that two-kink and four-kink states do not show up in the topological sector).

\begin{figure}[t] \centering
\includegraphics[width=0.99\columnwidth]{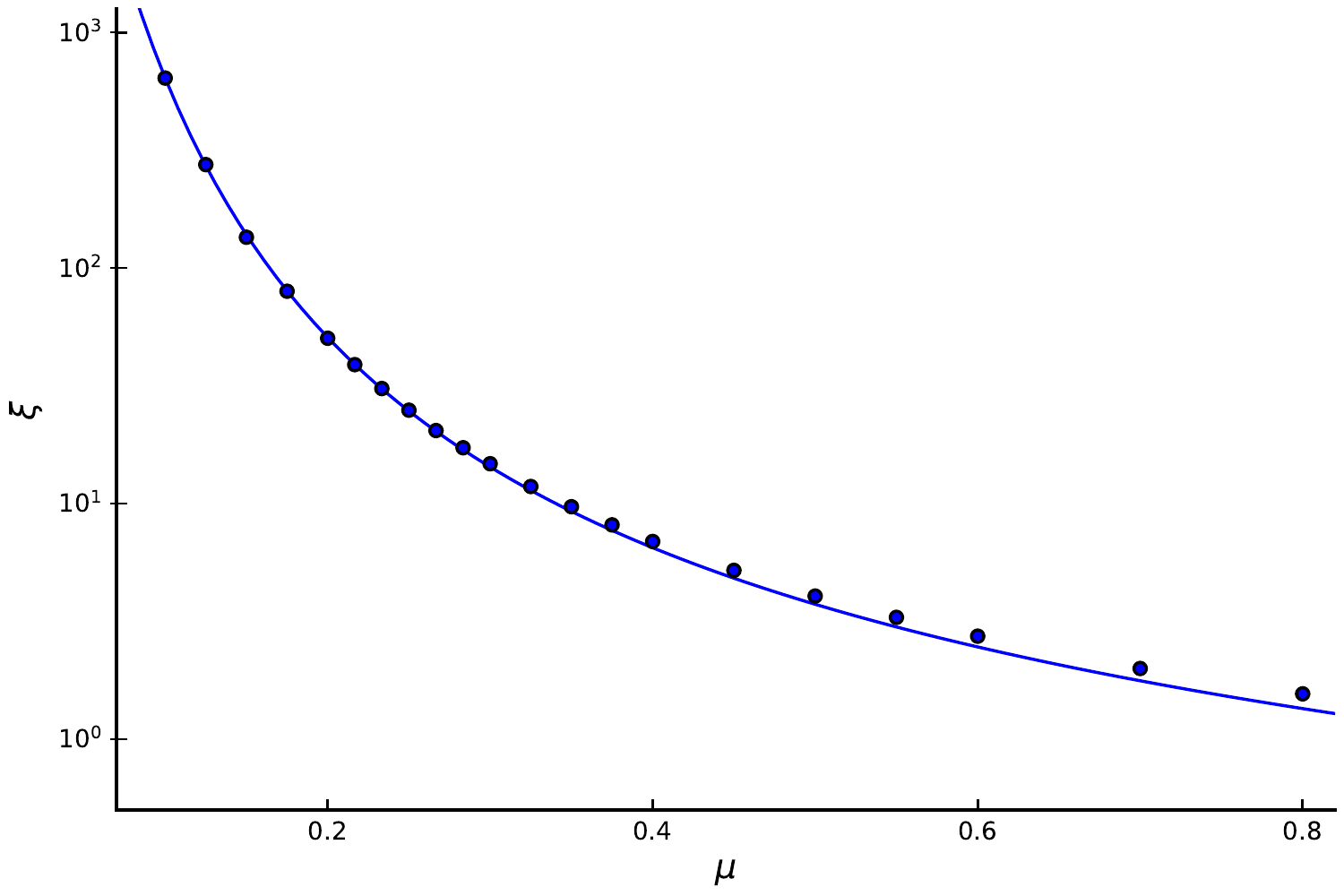}
\caption{The correlation length $\xi$ as a function of $\mu$ as the BKT transition is approached; the values for the correlation length were extrapolated from MPS simulations up to bond dimension $D=70$. We have fitted (blue line) these values to the form $\log\xi\propto(\mu-\mu_c)^{-1/2}$; we find a value of about $\mu_c\approx-0.05$, which is consistent with the expected value of $\mu_c=0$ given the relatively small bond dimensions used in our simulations.}
\label{fig:kt}
\end{figure}

\par
\begin{figure}[t] \centering
\includegraphics[width=0.99\columnwidth]{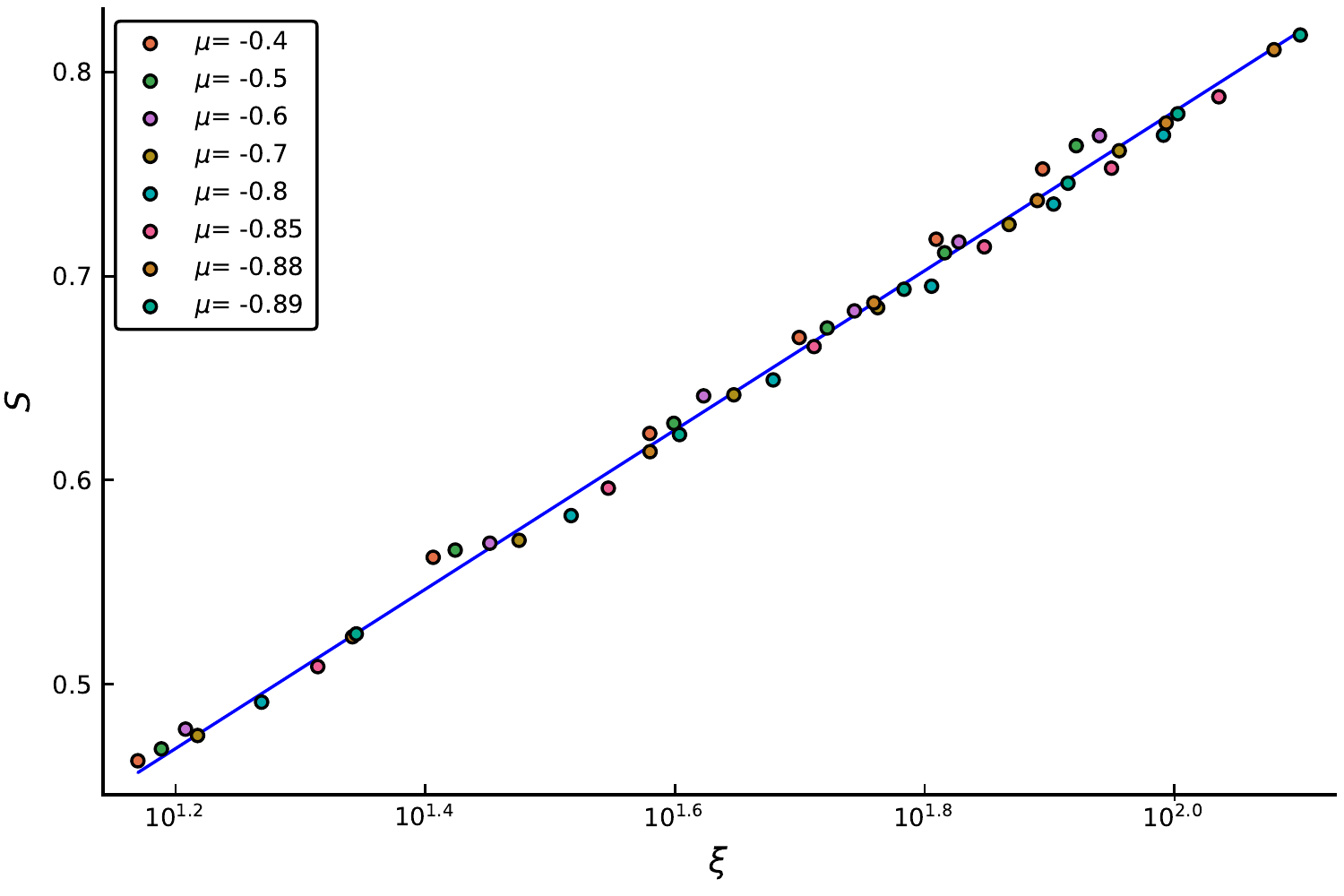}
\caption{The bipartite entanglement entropy $S$ as a function of the correlation length $\xi$ for variational MPS ground states at different values of $\mu$ in the critical region and varying bond dimensions. In the gapless phase, we consistently find a scaling $S = \frac{c}{6}\log\xi + S_0$ with values of $c$ around 1, where $S_0$ is a non-universal (i.e. $\mu$-dependent) constant related to the UV-scale of the problem. To collapse the data for different values of $\mu$, we have subtracted $S_0$. Fitting a single line through all data points, we find $c\approx1.02$ (blue line), in very good agreement with the effective Luttinger-liquid theory.}
\label{fig:central}
\end{figure}

 From the excitation spectra in the topological sector (kink sector), we learn that the kink gap closes around $\mu=0$, resulting in the condensation of kink-antikink pairs. Because of the non-trivial fusion properties of these kinks, the result cannot be an isolated critical point. Rather, the model enters a gapless phase for $\mu \leq 0$. The central charge in this phase can be determined from MPS simulations through the scaling of the entanglement entropy as a function of the effective correlation length in the MPS ground-state approximations (a technique knowns as finite-entanglement scaling \cite{Tagliacozzo,PollmannMukerjee}). In Fig.~\ref{fig:central} we clearly show that the central charge is $c=1$ throughout the gapless phase. The phase transition from this $\mathrm{U}(1)$ phase into the gapped phase for $\mu>0$ is expected to be of the BKT type, which we can confirm from the behavior of the correlation length as the critical point is approached in the gapped phase. The correlation length is a quantity that converges slowly with the bond dimension, so we apply extrapolation techniques \cite{Marek} for finding the correct value of the correlation length at each value of $\mu > 0$. In Fig.~\ref{fig:kt} we observe that the correlation length diverges exponentially as $\log\xi \propto \left(\mu-\mu_c\right)^{-1/2}$, and find a value for the critical point that is close to $\mu_c=0$.

\par For large negative $\mu$ we expect to recover the properties of the antiferromagnetic Ising model, for which the order parameter is the staggered magnetization $\braket{(-1)^i\sigma^z_i}$. One can see that the staggered magnetization clearly signals the phase transition into the gapless phase around $\mu=-0.9$, where it drops discontinuously to zero. This suggests that the transition is first order, which is confirmed by plotting the behavior of the correlation length as a function of the bond dimension in our MPS simulations (inset). We observe that the correlation length remains finite upon approaching the transition from the antiferromagnetic side. In fact, even arbitrarily close to the first-order transition, the correlation length in the antiferromagnetic phase remains of order one.

\begin{figure}[t] \centering
\includegraphics[width=0.99\columnwidth]{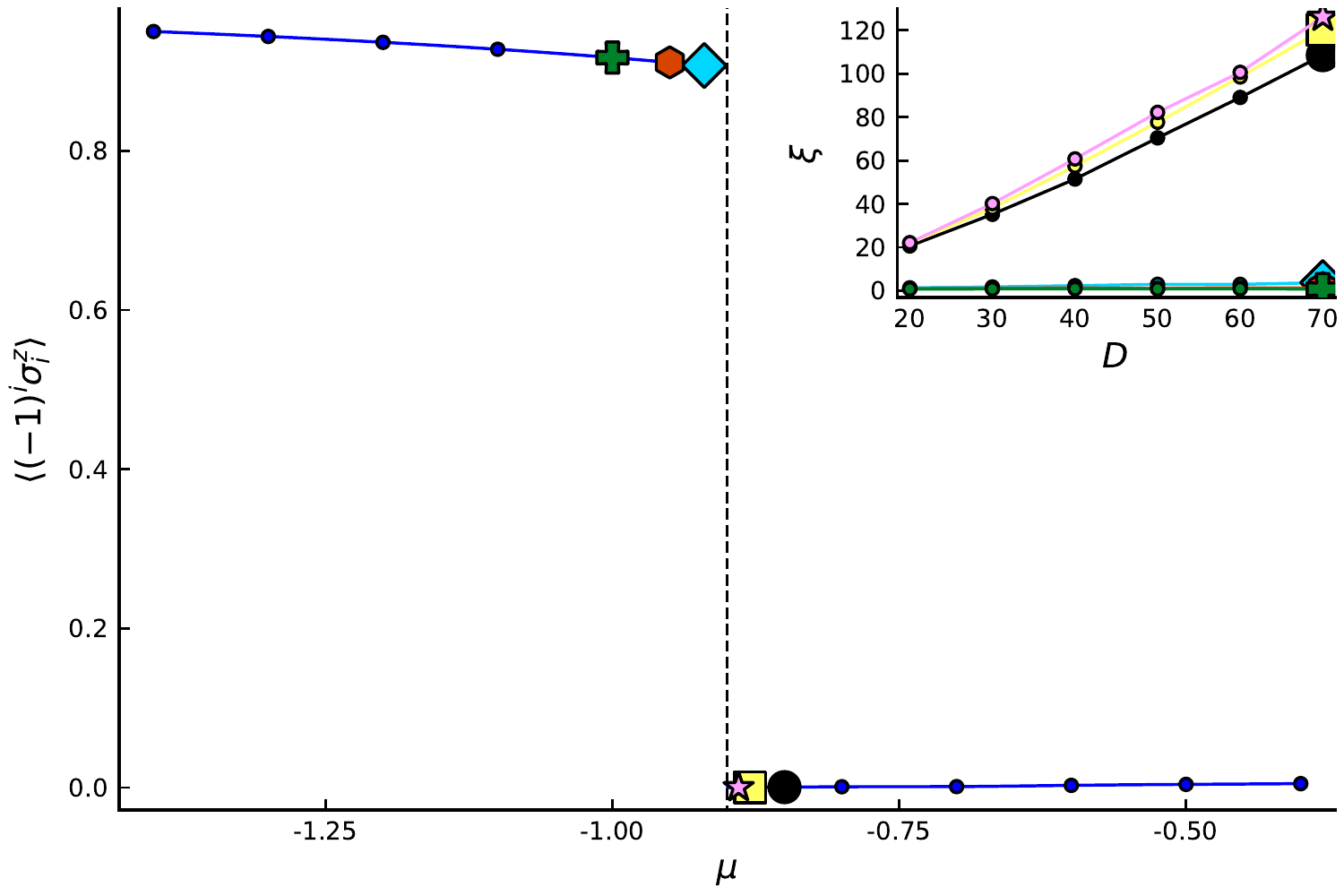}
\caption{The staggered magnetization $m_s=\braket{(-1)^i \sigma^z_i}$ as a function of $\mu$; the discontinuous jump clearly signals a first-order transition. As a further confirmation, in the inset we plot the effective correlation lengths as a function of the MPS bond dimension $D$ for different values of $\mu$ around the transition (the markers indicate the chosen values); in the critical region the correlation length grows indefinitely, whereas in the gapped region it remains clearly finite.}
\label{fig:first}
\end{figure}

\section{Discussion}

One-dimensional Hamiltonians with an anomalous $\mathbb{Z}_2$ MPO symmetry have previously been studied in the literature \cite{CZX,SantosWang,Bridgeman2,Williamson}, but to the best of our knowledge these models do not contain a parameter that enables a spontaneous breaking of the anomalous symmetry, and are therefore not based on a physical picture of the unconventional domain wall fusion properties. Ref. \cite{JuvenWang} did construct microscopic models to study domain walls on the boundaries of SPT states, but only those with fractional quantum numbers or non-trivial projective representations. The modified Ising chain is expected to be closely related to strange correlator partition functions obtained from the non-trivial bosonic $\mathbb{Z}_2$ SPT phase, as studied previously in Refs. \cite{StrangeCorr,Scaffidi,Bultinck,GuanMing}. These strange correlator partition functions have a natural interpretation as loop models \cite{StrangeCorr,Scaffidi}, which makes them tractable for certain analytical calculations. However, there is no unique strange correlator partition function, and different partition functions can give rise to very different critical behavior. Given the simplicity and associated physical picture of the modified Ising chain, we expect it to be a faithful effective model for the boundary of the non-trivial $\mathbb{Z}_2$ SPT. It would therefore be interesting to understand whether the modified Ising chain can be mapped to other (integrable) models known in the literature. Because of its simplicity it might also be worth to see if it can be realized in cold atom experiments.

The phase diagram we obtain for the modified Ising chain is very similar to that of the spin-$1/2$ XXZ model. Upon varying the chemical potential for the domain walls, we find three phases: a ferromagnetic phase, a Luttinger liquid regime and an antiferromagnetic phase. The ferromagnetic phase is separated from the Luttinger liquid by a BKT transition and the antiferromagnetic phase is separated from the Luttinger liquid by a first-order transition. In the antiferromagnetic regime, the correlation length in our model stays order of order one, even close to the first order transition into the Luttinger liquid. This behavior is again very similar to the XXZ spin chain, where the correlation length is exactly zero in this regime. Interestingly, the XXZ Hamiltonian also has a `perturbative' anomaly that is closely related to the `t Hooft anomaly of our model \cite{Ryu2,Furuya,MetlitskiThorngren}. The perturbative anomaly in the XXZ model is associated to translation symmetry, which acts on the low-energy modes as an effective $\mathbb{Z}_2$ symmetry. This anomaly is the same as the chiral anomaly of the 1D Dirac fermion \cite{Ryu}, and the $g\leftrightarrow -g$ anomaly of the SU$(2)_1$ Wess-Zumino-Witten CFT \cite{Gepner,Furuya}. It is also the same anomaly as the one associated with the non-local MPO symmetry of our model \cite{Furuya,MetlitskiThorngren,Bultinck}.

Condensation of defects in the order parameter of spontaneously broken anomalous symmetries often leads to emergent symmetries. In fact, such emergent symmetries are one of the hallmarks of deconfined quantum critical points \cite{DQCP,Nahum2,Nahum}. For example, at the proposed deconfined quantum critical point describing the 2D N\'{e}el-VBS transition, there is an emergent SO$(5)$ symmetry which allows to rotate between the N\'{e}el and VBS order parameters, which should be treated on equal footing at the transition point \cite{Nahum2}. In the 1D model studied here, there is a U$(1)\times$U$(1)$ symmetry that emerges after condensing the anomalous domain walls, although there is no deconfined quantum critical point or a physical interpretation for the emergent symmetry in terms of rotating between different order parameters. One point to make in this context is that the emergent U$(1)\times$U$(1)$ symmetry of the Luttinger liquid, acting as $(\phi,\theta) \rightarrow (\phi+\alpha_1,\theta+\alpha_2)$, is closely related to the self-duality of the free boson CFT, which interchanges $\phi$ and $\theta$. From the symmetry action \eqref{symmaction}, we see that both $\phi$ and $\theta$ can serve as order parameters for the $\mathbb{Z}_2$ symmetry, so it is the self-duality which states the equivalence of these two order parameters (which signal the breaking of the \emph{same} symmetry) at the BKT transition. In the context of deconfined quantum critical points, recent progress has shown that also in two spatial dimensions emergent symmetries are often closely related to dualities, in the sense that knowing dual formulations of a particular theory can help in understanding its emergent symmetries \cite{Nahum}.

Our construction of the microscopic model can be generalized to arbitrary discrete groups, by doing a similar `anyonic chain' construction with the group cohomology data instead of the $F$-symbols of a modular category. As was shown in Ref. \cite{LuLee} using a field theory analysis, in the case of $\mathbb{Z}_N$ symmetry, the domain walls are expected to have parafermionic statistics \cite{FradkinKadanoff}. It would be interesting to understand the connection between these symmetry-broken phases and the recently studied parafermionic chains \cite{Fendley,AliceaFendley}, which were argued to also realize BKT transitions \cite{Cheng}. The anyonic chain construction, however, is not restricted to Abelian symmetries and realizes domain walls which cannot be captured by the parafermion formalism. In particular, we can even go back to the original anyonic chain construction based on $F$-symbols and ask what is the precise nature of the defects in the corresponding `symmetry-broken' phases of the non-local MPO `symmetries' \cite{Trebst,Trebst2}. These `symmetry-broken' phases are closely related to gapped boundaries of 2D topologically ordered phases. This connection is manifested clearly in tensor network representations of the relevant topological phases, and has been exploited to study anyon condensation numerically \cite{Shadows,Marien2017,iqbal2018study}.

\section{Conclusion}
In this work, we have constructed a simple spin chain Hamiltonian that exhibits an anomalous $\mathbb{Z}_2$ symmetry, by explicitly modelling the semion statistics of the associated $\mathbb{Z}_2$ domain wall configurations. The resulting Hamiltonian is analogous to the Ising model, and shares with it an ordered ferromagnetic phase and antiferromagnetic phase. However, due to the anomalous realization of the symmetry, a gapped disordered phase is ruled out \cite{CZX}. Instead, we find a gapless Luttinger liquid phase that is separated from the ordered ferromagnetic phase by a phase transition of the BKT type, and from the ordered antiferromagnetic phase by a first order phase transition. This model is believed to capture the universal physics of the boundary of 2D SPT phases.

The reasoning on which this work is based, can also be applied to the boundaries of 3D SPT phases with discrete symmetries. In the symmetry-broken phase, the `t Hooft anomaly will manifest itself via unconventional properties of the junctions of domain walls, which again have a natural interpretation in terms of the group cohomology data specifying the SPT phase. If one could construct an effective model that captures the anomalous properties of the domain wall junctions, then one could perhaps gain some insight in the boundary phase diagram of 3D SPT phases. 

\emph{Acknowledgements--} We thank Frank Verstraete for helpful and stimulating discussions, and an anonymous referee for helping us clarify our arguments in section IV. NB acknowledges an inspiring discussion with Ruben Verresen. GR and JH are supported by the European Research Counsil (ERQUAF 715861), LV by the Research Foundation Flanders and NB by a BAEF Francqui Fellowship.

\bibliography{bibliography}

\end{document}